\documentclass[copyright,creativecommons]{eptcs}
\providecommand{\event}{QPL 2013}
\usepackage{amsmath,amsthm,amssymb,xspace}
\usepackage[all]{xy}

\usepackage{graphicx}
\usepackage{amsmath, amsthm, amssymb}
\usepackage{subfigure}
\usepackage{comment}
\usepackage{bm}
\usepackage{epsfig,color}
\usepackage[sans]{dsfont}

\newcommand{\be}{\begin{equation}}
\newcommand{\ee}{\end{equation}}
\newcommand{\bea}{\begin{eqnarray}}
\newcommand{\eea}{\end{eqnarray}}
\newcommand{\bes}{\begin{equation*}}
\newcommand{\ees}{\end{equation*}}
\newcommand{\beas}{\begin{eqnarray*}}
\newcommand{\eeas}{\end{eqnarray*}}

\newcommand{\x}{\mathrm{x}}

\def\x{\mathrm{x}}
\def\y{\mathrm{y}}
\def\g{\mathrm{guess}}

\def\x{\mathrm{x}}
\def\y{\mathrm{y}}
\def\w{\mathrm{w}}

\newtheorem*{thm*}{Theorem}

\newtheorem*{lem*}{Lemma}

\newtheorem*{lipschitzLem*}{Lemma \ref{lipschitz}}
\newtheorem*{lipschitzCubeLem*}{Lemma \ref{lipschitzCube}}
\newtheorem*{pgmNearlyOptimalThm*}{Theorem \ref{pgmNearlyOptimal}}

\title{ Distinguishability, Ensemble Steering, and the No-Signaling Principle }

\author{Joonwoo Bae  \\ \\[0.3em]
{\it\small  Center for Quantum Technologies, National University of Singapore, Singapore 117543,
}\\
{\it\small  ICFO--Institut de Ci\`encies Fot\`oniques, 08860 Castelldefels (Barcelona), Spain.}}

\def\titlerunning{Distinguishability, Ensemble Steering, and the No-Signaling Principle}
\def\authorrunning{Joonwoo Bae }

\begin{document}

\maketitle

\begin{abstract}
We consider a fundamental operational task: distinguishing systems in different states in the framework of generalized probabilistic theories. We provide a general formalism of minimum-error discrimination of states in convex optimization. With the formalism established, we show that the distinguishability is generally a global property assigned to the ensemble of given states rather than other details of a given state space or pairwise relations of given states. Then, we consider bipartite systems where ensemble steering is possible, and show that show that with two operational tasks (ensemble steering and satisfying the no-signaling condition) the distinguishability is tightly determined. The result is independent of the structure of the state space. This concludes that the distinguishability is generally determined by the compatibility between two tasks, ensemble steering of states and the no-signaling principle. 
\end{abstract}



\section{Introduction}

One of the most fundamental tasks of information processing is to distinguish different signals that deliver distinct messages. If quantum systems are used to carry messages from one party to another, we are led to the problem of discriminating between different quantum states. In a naive way, if messages are encoded in distinguishable quantum states i.e. orthogonal states, which are then transmitted through a noisy channel, during which the quantum states may interact with the environment resulting in a state that may no longer be orthogonal. Non-orthogonal quantum states cannot be perfectly distinguished in a single-shot manner, that is, they are indistinguishable. Then, for the practical information application of quantum systems, one has to devise methods of quantum state discrimination and optimization of measurement for these purposes. 

Interestingly, the indistinguishability of non-orthogonal quantum states seems to have a deeper meaning than merely a problem that can be considered for practical applications. While indistinguishability itself has appeared in contexts of classical information processing through probabilistic systems which cannot be distinguished perfectly, recent works have shown that it cannot be consistently reproduced by probabilistic classical systems only \cite{ref:leifer}. Indistinguishability of quantum states has been indeed a fundamental reason behind many quantum information applications. For information-theoretic purposes, it is useful to have a direct characterization of the distinguishability in an operational way, however, there is little known so far in this direction. 

In this work, we provide an operational characterization of the indistinguishability of quantum states in terms of two operational tasks, ensemble steering and the no-signaling principle. We show that a physical theory in general, where a system is described as a box producing probabilistic outcomes, indistinguishability is tightly connected to the competition between two tasks: one is ensemble steering on sources i.e. the ability of manipulating states of systems from a distance in a remote way before measurement happens on systems; and the other is the no-signaling constraint on measurement outcomes. In other words, distinguishability cannot be improved even better if ensemble steering is allowed and the no-signaling principle is respected. Then, quantum theory is merely an instance of the case; the quantum indistinguishability can be characterized by ensemble steering on quantum states and the no-signaling principle on measurement outcomes.

\section{Preliminaries}

Before we proceed to the main and technical details, let us set notations and basic definitions. We begin with the framework in GPTs where notions of states, measurement, and their relations are generalized \cite{ref:birvon, ref:barnum1, ref:barrett}. Here we will make use of the mathematical framework in Ref. \cite{ref:barnum1}. 

In a GPT, the set of states, denoted by $\Omega$, consists of all possible states that a system can be prepared in. Any probabilistic mixture of states, i.e. $pw_1 + (1-p) w_2 \in \Omega$ for $w_{1},w_{2}\in \Omega$ and probability $p$ is also a state, i.e. $\Omega$ is convex. A general mapping from states to probabilities is called {\it effects} and described by linear functionals $\Omega\rightarrow[0,1]$. A measurement denoted by $s$ corresponds to set of effects, $E^{(s)} = \{e_{\x}^{(s)} \}_{\x=1}^{N}$, by which the probability of getting outcome $\x$ given a state $w$ is, $p(\x|s) = e_{\x}^{(s)}[w]$. A unit effect $u$ means a measurement that occurs, that is, $u[w]=1$ for all $w\in \Omega$, so that  for any measurement $s$, it holds $\sum_{\x} e_{\x}^{(s)} = u$. As effects are dual to the state space, they are also convex. 

Here, we identify state distinguishability with minimal error, or equivalently, the maximal success probability in a guessing task for different states. This is called minimum-error state discrimination, and can be described by a game with two parties, Alice and Bob, as follows. Suppose there is a set of states agreed by them in advance, and Alice prepares a system in one of $N$ states with some probability and gives it to Bob. If Bob makes a correct guess, their score is given $1$, otherwise $0$. Their goal is to maximize the average score over all measurements. We write the states by $\{ w_{\x}\}_{\x=1}^{N}$ and Alice's {\it a priori} probabilities by $\{ q_{\x}\}_{\x=1}^{N}$, and then by $\{q_{\x}, w_{\x} \}_{\x=1}^{N}$ altogether. Bob has to find optimal measurement to maximize the score. We write by $p_{B|A} (\x | \y) = e_{\x} [w_{\y}]$ the probability that Bob makes a guess $w_{\x}$ when state $w_{\y}$ is given by Alice. Now, the goal is to find the maximal success probability of making a correct guess, the so-called \emph{guessing probability} in the following,
\bea
p_{\g} :=  \max  \sum_{\x=1}^{N}  q_{\x} p_{B|A}(\x | \x)=   \max_{\{ e_{\x}\}_{\x=1}^{N}}   \sum_{\x=1}^{N} q_{\x} e_{\x} [w_{\x}] \label{eq:gp}
\eea
where the maximization runs over all effects. Note that GPTs are generally not self-dual, meaning an isomorphism between two spaces does not exist in general \cite{ref:brunner}. 

\section{Indistinguishability via optimal discrimination}

Since state and effect spaces are convex, it is possible to formalize the optimization task in Eq. (\ref{eq:gp}) in the convex optimization framework \cite{ref:boyd}. For states $\{q_{\x},w_{\x} \}_{\x=1}^{N}$, we take the form in Eq. (\ref{eq:gp}) as the primal problem denoted by $p^{*}$
\bea
p^{*} & = & \max\{ \sum_{\x=1}^{N} q_{\x} e_{\x} [ w_{\x} ] ~\|~ e_{\x}\geq0 ~\forall\x, ~\sum_{\x=1}^{N} e_{\x} = u  \}   \label{eq:primal} 
\eea 
and derive its dual $d^{*}$, as follows 
\bea
d^{*} & = & \min \{ u [K ]~\| ~K\geq q_{\x} w_{\x},~\x=1,\cdots,N \} \label{eq:dual}
\eea
where inequalities mean the order relation in the convex set: by $e_{\x}\geq 0$, it is meant that $e_{\x}[w] \geq 0$ for all $w\in \Omega$, and by $K\geq q_{\x} w_{\w}$, that $e[K-q_{\x} w_{\x} ]\geq 0$ for all effects $e$. Following from the so-called Slater's constraint quantification in convex optimization, a sufficient condition for the strong duality is the strict feasibility, which is the existence of a strictly feasible point of parameters. For instance, primal parameters $\{e_{\x} = u/N\}_{\x=1}^{N}$ are in the case, since $e_{\x} [w_{\y}]>0$ $\forall \x, \y$ and $\sum_{\x} e_{\x}=1$. Thus, the strong duality holds true, meaning that both solutions from primal and dual problems are equal, i.e. $p^{*} = d^{*}$. 


There is another approach called the \emph{complementarity problem} that generalizes convex optimization, in which the optimality conditions of a given optimization problem are directly analyzed. This approach has been applied to quantum state discrimination in the context of two-state discrimination in Refs. \cite{ref:hwang, ref:baehwang}. The approach has been cited and generalized to GPTs in Ref. \cite{ref:kimura} under the assumption that conjugate states, equivalent to complementary states in the below, do exist. There those classes of quantum states for which conjugate states exist are named the Helstrom family, the existence of which has remained in question. Therefore, the results are valid only under the unproven assumption. In fact, a formal statement of the complementary problem for optimal state discrimination in GPTs, that is, a formal way that provides solutions in an explicit way, has been shown in Ref. \cite{ref:baearxiv}. In what follows, we provide the complementary problem in a formal way via convex optimization, through which the existence of all parameters is immediately guaranteed. 

The complementarity problem deals with both primal and dual parameters in Eqs. (\ref{eq:primal}) and (\ref{eq:dual}) which in general is not more efficient that the convex optimization approach. The advantage, however, lies at the fact that generic structures existing in the problem are exploited. The optimality conditions can be summarized by the so-called Karush-Kuhn-Tucker (KKT) conditions, which are constraints listed in Eqs. (\ref{eq:primal}) and (\ref{eq:dual}), together with the followings,
\bea
\mathrm{ (Symmetry~parameter) }&&~ K = q_{\x} w_{\x} + r_{\x} d_{\x} ,~\forall ~\x \label{eq:kkt1} \\
\mathrm{(Orthogonality)}&&~ e_{\x} [r_{\x} d_{\x}]  = 0, ~\forall~\x, \label{eq:kkt2}
\eea
where $r_{\x} \in [0,1]$ for all $\x$, and $\{ d_{\x}\}_{\x=1}^{N}$ which we call complementary states are normalized, i.e. $u [ d_{\x} ]=1$. The first condition, the symmetry parameter, follows from the Lagrangian stability and shows that for any discrimination problem e.g. $\{ q_{\x}, w_{\x} \}_{\x=1}^{N}$, there exists a single parameter $K$ which is decomposed into $N$ different ways with given states and complementary states $\{r_{\x}, d_{\x}\}_{\x=1}^{N}$. Then, the second condition in Eq. (\ref{eq:kkt2}) from the complementary slackness characterizes optimal effects by the orthogonality relation between complementary states and optimal effects. These  generalize optimality conditions for quantum states in Refs. \cite{ref:holevo} \cite{ref:yuen} to all GPTs, see also different forms of optimality conditions \cite{ref:baegeo}. Primal and dual parameters satisfying the KKT conditions are automatically optimal parameters that provide solutions to optimal discrimination. Moreover, for the problem here, recall that the strong duality holds i.e. $p^{*} = d^{*}$. Conversely, the fact that the strong duality holds in Eqs. (\ref{eq:primal}) and (\ref{eq:dual}) implies the existence of optimal parameters which satisfy KKT conditions and give the guessing probability in Eq. (\ref{eq:gp}). Note that we derive all these from the fact that state spaces of GPTs are convex. 

The complementarity problem can then be formalized in a geometric way. We first remark that, in optimality conditions in Eqs. (\ref{eq:kkt1}) and (\ref{eq:kkt2}), constraints for states and effects are separated. The symmetry parameter $K$ is characterized on a state space and gives the guessing probability, see Eq. (\ref{eq:dual}), i.e. $p_{\g} = u [K ] = q_{\x} + r_{\x}$. Then, the guessing probability can be found by searching complementary states $\{r_{\x},d_{\x} \}_{\x=1}^{N}$ fulfilling Eq. (\ref{eq:kkt1}) on the state space. This can be described in a systematic way, as follows. Let us define a polytope denoted by $\mathcal{P}(\{ q_{\x}, w_{\x} \}_{\x=1}^{N})$ of given states in the state space: each vertex of the polytope corresponds to unnormalized state $q_{\x} w_{\x}$ for $\x=1,\cdots,N$. Then, the polytope of complementary states, $\mathcal{P}(\{ r_{\x}, d_{\x} \}_{\x=1}^{N})$, is in fact immediately congruent to $\mathcal{P}(\{ q_{\x}, w_{\x} \}_{\x=1}^{N})$ in the state space, since the following holds from Eq. (\ref{eq:kkt1}),
\bea
q_{\x} w_{\x} - q_{\y} w_{\y} = r_{\y} d_{\y} - r_{\x} d_{\x},~~\mathrm{for~all}~\x,\y, \label{eq:kkt}
\eea
which shows that corresponding lines of two polytopes $\mathcal{P}(\{ q_{\x}, w_{\x} \}_{x=1}^{N})$ and $\mathcal{P}(\{ r_{\x}, d_{\x} \}_{x=1}^{N})$ are of equal lengths and anti-parallel. Then, from the underlying geometry of the state space, one can find complementary states by putting two congruent polytopes such that the condition in Eq. (\ref{eq:kkt1}) holds. Optimal effects can be found from the orthogonal relation in Eq. (\ref{eq:kkt2}), accordingly. 

For {\it a priori} probabilities given as $q_{\x} = 1/N$, the guessing probability becomes even simpler. First, it follows $r_{\x} = r_{\y}$ for all $\x,\y$: this is obtained from the expression $p_{\g} = q_{\x} +r_{\x}$ for any $\x$, see Eqs. (\ref{eq:kkt1}) and (\ref{eq:dual}). Denoted by $r:=r_{\x}$ for all $\x$, the guessing probability is now,
\bea 
p_{\g} = \frac{1}{N} + r,~~\mathrm{with}~~r = \frac{ \| \frac{1}{N} w_{\x} - \frac{1}{N} w_{\y}  \| }{ \|  d_{\x} - d_{\y}  \|}\label{eq:r}
\eea
where the expression of $r$ follows from the condition in Eq. (\ref{eq:kkt}) with a distance measure $\| \cdot\|$ that can be defined in the state space. The parameter $r$ has a meaning as the ratio between two polytopes, $\mathcal{P}(\{ 1/N, w_{\x} \}_{x=1}^{N})$ of given states, and  $\mathcal{P}(\{d_{\x} \}_{x=1}^{N})$ of only complementary states.

\section{Tightness}

We now move to a bipartite scenario where the state space of two parties is constrained such that ensemble steering is possible. This means that Alice can steer to any decomposition of Bob's ensemble. In quantum theory, the notion of steering was introduced by Schr\"odinger \cite{ref:schr} and then, with specification to a bipartite Hilbert space, formalized by the so-called Gisin-Hughston-Jozsa-Wootters theorem \cite{ref:ghjw1,ref:ghjw2}. Note that ensemble steering does not yet single out quantum theory among GPTs \cite{ref:barnum2}. We also distinguish the extension from the purification lemma which fully characterizes quantum theory \cite{ref:guilio}.

In what follows, we apply the theoretical tools developed so far, and show that for any GPTs endowed with ensemble steering, the distinguishability is immediately determined by a way of excluding instantaneous communication. We first derive a bound to the optimal distinguishability in a given GPT by the no-signaling condition, and then we prove that the bound is tight, i.e. it can be achieved within the given GPT. The result is independent of particular properties of a state space. 

Let us incorporate state discrimination to the following non-signaling framework. Let $\{q_{\x},w_{\x} \}_{\x=1}^{N}$ denote the states among which we are interested in discriminating. Suppose Alice steers the ensemble of Bob, denoted by $w_{B}$, in $N$ different decompositions. That is, the ensemble has $N$ different decomposition as $w_{B} = w_{B}^{(\x)}$ for $\x=1,\cdots,N$ where
\bea 
w_{B}^{(\x)} = p_{\x} w_{\x} + (1-p_{\x}) c_{\x}, ~\mathrm{with~}q_{\x} = \frac{p_{\x}}{\sum_{\x'=1}^{N} p_{\x'}  } ~ \label{eq:ens}
\eea
with some states $\{c_{\x}\}_{\x=1}^{N}$ and probabilities $\{p_{\x}\}_{\x=1}^{N}$. By ensemble steering, we mean that any of the $N$ decompositions of Bob's ensemble can be prepared by Alice's steering. Since Bob holds an identical ensemble, his measurement gains no knowledge about which decomposition is given, until Alice announces about her steering. The non-signaling condition is thus naturally imposed.

The distinguishability on $\{q_{\x},w_{\x} \}_{\x=1}^{N}$ is then constrained by the non-signaling condition as follows. Assume that Bob optimizes measurement to guess which state among $\{w_{\x} \}_{\x=1}^{N}$ exists in his ensemble. The strategy is, once the state $w_{\x}$ is found, he concludes his ensemble is in the decomposition $w_{B}^{(\x)}$, see Eq. (\ref{eq:ens}), by which he also guesses Alice's steering strategy. Then, by the no-signaling condition, discrimination among states $\{w_{\x} \}_{\x=1}^{N}$ must be constrained so that Bob would not learn Alice's steering better than the random guess. 

We now derive an upper bound to the guessing probability by the no-signaling condition. Let $P_{B|A} (\x | \y)$ denote the probability that, while Alice has actually steered ensemble $w_{B}^{( \y )}$, Bob concludes his ensemble in $w_{B}^{(\x)}$ by discriminating among $\{q_{\x},w_{\x} \}_{\x=1}^{N}$. The no-signaling condition \cite{ref:barrett} \cite{ref:masanes} implies the following constraint
\bea 
\sum_{\x=1}^{N} P_{B|A} (\x | \x) \leq 1. \label{eq:nosig} 
\eea
If the condition is not fulfilled, one can explicitly construct a superluminal communication protocol \cite{ref:bae}. Then, recall Bob's strategy of guessing Alice's steering: to guess Alice's steering strategy, he attempts to distinguish ensemble decompositions $\{w_{B}^{(x)} \}_{\x=1}^{N}$ by exploiting optimal discrimination of states $\{w_{\x}\}_{\x=1}^{N}$ existing in the ensemble.

If Alice has steered Bob's ensemble $w_{B}^{(\x)}$, Bob's correct conclusion happens when i) $w_{\x}$ is given, which appears with probability $p_{\x}$, and ii) measurement gives a correct answer, that is, with probability $p_{\x} p_{B|A}(\x |\x)$. In the strategy, there can be contribution in measurement from the other state $c_{\x}$ in the ensemble with probability $1-p_{\x}$. Thus, it holds, $p_{\x} p_{B|A}(\x |\x) \leq P_{B|A}(\x | \x)$. In addition, recall that measurement is optimized for discrimination among $\{q_{\x}, w_{\x}\}_{\x=1}^{N}$, since the \emph{a priori} probability for state $w_{\x}$ among $\{w_{\x}\}_{\x=1}^{N}$ is given by $q_{\x}$, see Eq. (\ref{eq:ens}). From the no-signaling condition in Eq. (\ref{eq:nosig}), we have $\sum_{\x=1}^{N} p_{\x}p_{B|A}(\x |\x)\leq1$, from which we have
\bea
p_{\g} = \max\sum_{\x} q_{\x} p_{B|A} (\x| \x) \leq \frac{1}{p_{1} + \cdots + p_{N}}. \label{eq:gpns}
\eea
Thus, a upper bound to the distinguishability is obtained from the no-signaling condition, and expressed in terms of parameters $\{ p_{\x}\}_{\x=1}^{N}$ of steering each state in $\{ w_{\x}\}_{\x=1}^{N}$.

We then show that the bound is indeed tight, i.e. it can be achieved within a given GPT. We show the tightness by proving that, for any set $\{q_{\x},w_{\x} \}_{\x=1}^{N}$, the optimal discrimination characterized by the KKT conditions implies the existence of both an identical ensemble in Eq. (\ref{eq:ens}) and effects achieving the bound in Eq. (\ref{eq:gpns}).

Recall the general method of optimal discrimination, the existence of a symmetry parameter $K$ that completely characterizes the optimal distinguishability, see Eq. (\ref{eq:kkt1}). The parameter has $N$ decompositions with complementary states $\{r_{\x},d_{\x}\}_{\x=1}^{N}$. Its normalization $\widetilde{K} = K / u[K]$ shows, for each $\x$,
\bea
\widetilde{K}  = p_{\x} w_{\x} + (1-p_{\x}) d_{\x}, \mathrm{~with}~ p_{\x} = q_{\x}/u[K].\label{eq:tilK} 
\eea
This corresponds to ensemble steering in Eq. (\ref{eq:ens}). Recall the dual problem in Eq. (\ref{eq:dual}) which gives the guessing probability in GPTs, as $p_{\g} = u[K]$. Using the identity $\sum_{\x=1}^{N} q_{\x}=1$ and the relation $p_{\x} = q_{\x}/ u[K]$, the solution in the dual problem can be computed as, $u(K) = (\sum_{\x=1}^{N} p_{\x} )^{-1}$. This shows that the bound in Eq. (\ref{eq:gpns}) is already achieved within a given GPT, and hence the tightness is shown. In addition, optimal effects also exist with complementary states, see Eq. (\ref{eq:kkt2}). 

\section{Summary}

We have developed and established a general method of distinguishing states in GPTs. This generalizes i) the geometric formulation \cite{ref:baegeo1, ref:baegeo2} and ii) optimality conditions \cite{ref:holevo} \cite{ref:yuen} in the quantum case to GPTs. The formulation is also illustrated with an example (the four-state polygon system) a particularly interesting case where the bipartite extension shows the maximally non-local correlations \cite{ref:barrett} \cite{ref:brunner}. It is also shown that distinguishability and non-locality are independent resources. We also remark that in GPTs i) measurement for the optimal discrimination is generally not unique, and ii) sometimes no measurement give an optimal strategy, thus similar to the results in quantum cases in \cite{ref:baegeo1,ref:baegeo2} \cite{ref:helstrom} \cite{ref:hunter}. With the general formalism and tools developed, we have shown that for GPTs where ensemble steering is possible, the distinguishability can be determined by no-signaling condition. State istinguishability was also shown to be dictated by the relation between ensemble steering and the no-signaling condition. This also gives a physical motivation for why we have the Born rule in Hilbert space quantum mechanics as opposed to mathematical motivations (e.g. Gleason's theorem \cite{ref:gleason}): we can have both ensemble steering yet satisfy the no-signalling principle.

{\small
\bibliography{all}
}

\end{document}